# Gamma-Ray Spectral Variability of Cygnus X-1

M. L. McConnell[1], K. Bennett[2], H. Bloemen[3], W. Collmar[4], W. Hermsen[3], L. Kuiper[3], W. Paciesas[5], B. Phlips[6], J. Poutanen[7], J. M. Ryan[1], V. Schönfelder[4], H. Steinle[4], A. W. Strong[4], and A. A. Zdziarski[8]

[1]*Space Science Center, University of New Hampshire, Durham, NH 03824*
[2]*Space Science Department, ESTEC, Noordwijk, The Netherlands*
[3]*SRON – Utrecht, Utrecht, The Netherlands*
[4]*Max Planck Institute for Extraterrestrial Physics, Garching, Germany*
[5]*University of Alabama, Huntsville, AL*
[6]*George Mason University, Fairfax, VA*
[7]*Stockholm Observatory, SE-106 91 Stockholm, Sweden*
[8]*N. Copernicus Astronomical Center, Warsaw, Poland*

**Abstract.** We have used observations from CGRO to study the variation in the MeV emission of Cygnus X-1 between its low and high X-ray states. These data provide a measurement of the spectral variability above 1 MeV. The high state MeV spectrum is found to be much harder than that of the low state MeV spectrum. In particular, the power-law emission seen at hard X-ray energies in the high state spectrum (with a photon spectral index of 2.6) is found to extend out to at least 5 MeV, with no evidence for any cutoff. Here we present the data and describe our efforts to model both the low state and high state spectra using a hybrid thermal/nonthermal model in which the emission results from the Comptonization of an electron population that consists of both a thermal and nonthermal component.

## INTRODUCTION

It has long been recognized that the soft X-ray emission of Cygnus X-1 (~10 keV) generally varies between two discrete levels [1,2]. The 20–100 keV time history of Cygnus X-1, as derived from BATSE occultation data, is shown in the center panel of Figure 1. The top panel of Figure 1 shows the 20–100 keV power-law spectral index, as derived from the BATSE occultation data. These data cover most of the CGRO mission, from the launch in April of 1991 until the end of 1999. During the first few months of the CGRO mission (up until October of 1991), all-sky monitoring data from Ginga (1–20 keV) was available, confirming that the source was in its low X-ray state during this period [3]. From October of 1991 until December of 1995, there were only sporadic pointed X-ray observations of the soft X-ray flux from Cygnus X-1. It was not until the launch of RXTE, in December of 1995, that continuous data on the soft X-ray flux once again became available. The data from the RXTE All-Sky Monitor (ASM) are shown in the lower panel of Figure 1, in the form of the 2–10 keV count rate.

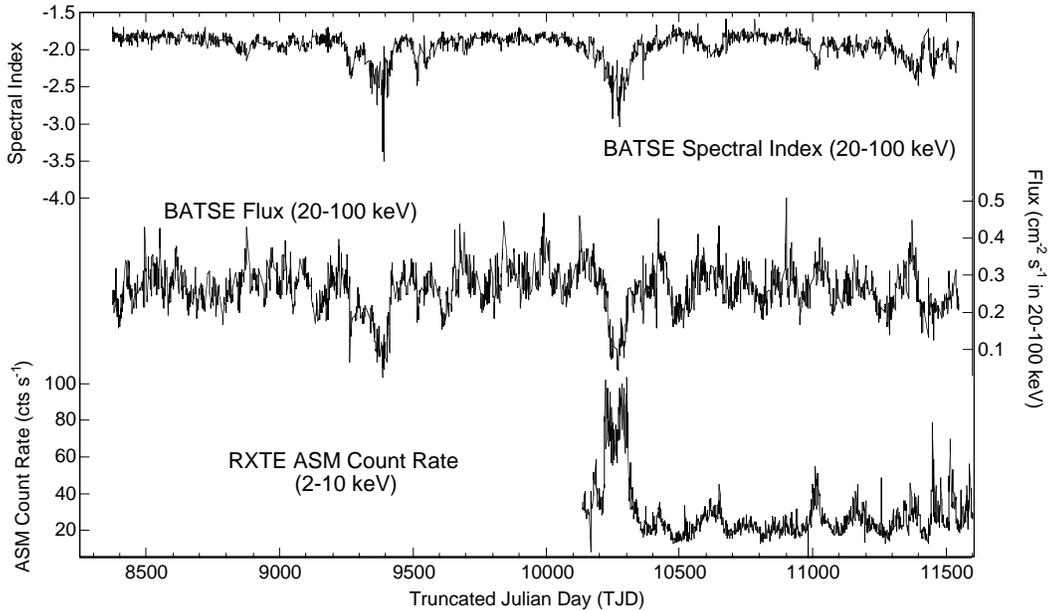

**FIGURE 1.** Time histories of the X-ray behavior of Cygnus X-1 for a time period encompassing most of the CGRO mission (1991–1999). The top two panels represent data from the BATSE experiment, showing the 20–100 keV flux (middle) and the power-law index for an assumed power-law spectrum in the 20–100 keV energy band (top). The lower panel shows the 2–10 keV count rate from the All-Sky Monitor (ASM) on RXTE.

The data shown in Figure 1 dramatically demonstrate the general X-ray behavior of Cygnus X-1. About 90% of the time, Cygnus X-1 is in its so-called "low" X-ray state (as defined by the soft X-ray emission). In this state, the soft X-ray flux (2–10 keV) is relatively low, while the hard X-ray flux (20–100 keV) is relatively high. The spectral shape in the 20–100 keV energy band is a relatively hard power-law spectrum with a photon spectral index near 1.8. During the so-called "high" X-ray state, the soft X-ray flux (2–10 keV) increases by about a factor of 3–4, while the hard X-ray flux (20–100 keV) decreases by about a factor of 3. This results in a much softer power-law spectrum, with a spectral index between 2.5 and 3.0. Data provided by all four instruments on CGRO provide an unprecedented opportunity to study the variability of the γ-ray emission between these two states.

## THE LOW STATE SPECTRUM

In earlier work [4], we assembled a broadband spectrum of the γ-ray emission from Cygnus X-1 by combining COMPTEL data with contemporaneous data from the BATSE [5,6], OSSE [7], and EGRET [8] experiments on CGRO. Due to the small FoV of OSSE (and the near all-sky monitoring capability of BATSE), the selection of data in this case was driven by the availability of contemporaneous OSSE data. The selection of data was also confined to the first three cycles of CGRO observations (1991–1994), based on the quality of the COMPTEL data that was available at the time the study was undertaken. A final selection on the data was imposed based on the level of hard X-ray flux and the spectral index observed by BATSE (Figure 1). This was done to ensure that there were no spectral state changes during the selected observations.

The general form of the spectrum (Figure 2) is that of the "breaking γ-ray state" that is generally associated with the low X-ray state of galactic black hole binaries [10]. The low-state spectrum provides evidence of significant emission out to at least 2 MeV. There is additional evidence for emission between 2–5 MeV, with no evidence for emission above 5 MeV. The data do not provide any evidence for a high-energy cutoff to the spectrum. Also shown in Figure 2 is an estimated upper limit from EGRET data collected during Cycles 1–4 of the CGRO mission, an exposure that is similar to that for the other data included in this analysis [8]. Unfortunately, the EGRET upper limit (based on an assumed $E^{-3}$ source spectrum) does not provide any further constraints on the extrapolated spectrum. Attempts were made to fit the broadband spectrum with both one- and two-component Compton models (based on [9]), and with an exponentiated power-law model. None of these models provided a statistically acceptable fit to the data. They all underestimated the measured flux levels at the highest energies near 1 MeV and above.

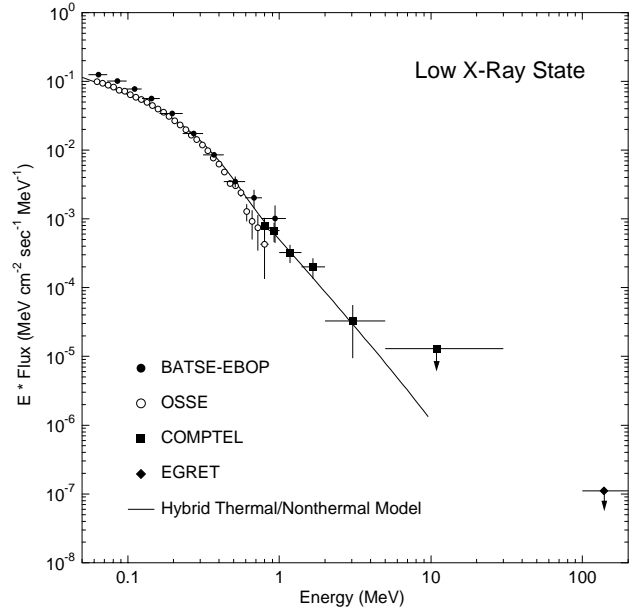

**FIGURE 2** - Contemporaneous broadband CGRO spectrum of the low X-ray state of Cygnus X-1. For the sake of clarity, upper limits from OSSE and BATSE are not shown, but these are consistent with the total dataset.

## THE HIGH STATE SPECTRUM

During the CGRO mission, Cygnus X-1 spent only about 10% of its time in the high X-ray state. The high X-ray state was clearly observed on only two occasions. In each case, the high state period lasted about 5 months. The high X-ray state was first observed by CGRO in January of 1994, at a time (prior to the launch of RXTE) when there was no soft X-ray monitoring data available. (This transition is clearly seen in Figure 1 near TJD 9400.) A CGRO target-of-opportunity was declared (CGRO viewing period 318.1) so that all four CGRO instruments (not just BATSE) could collect data. Observations by COMPTEL showed no detectable level of emission. This null result, however, was consistent with an extrapolation of the $E^{-2.7}$ power-law spectrum measured at hard X-ray energies by both BATSE [11] and OSSE [7].

The second observation of a high X-ray state took place in May of 1996. The transition was first observed by RXTE, beginning on May 10 [12]. The 2-12 keV flux reached a level of 2 Crab on May 19, four times higher than its normal value. Meanwhile, at hard X-ray energies (20-200 keV), BATSE measured a significant *decrease* in flux [13]. Motivated by these dramatic changes, a ToO for CGRO was declared and observations by OSSE and COMPTEL began on June 14 (CGRO viewing period 522.5). (Unfortunately, the EGRET experiment was turned off during this viewing period, as part of an effort to conserve its supply of spark chamber gas.) During the ToO, COMPTEL collected 11 days of data (from June 14 to June 25) at a favorable aspect angle of 5.3°.

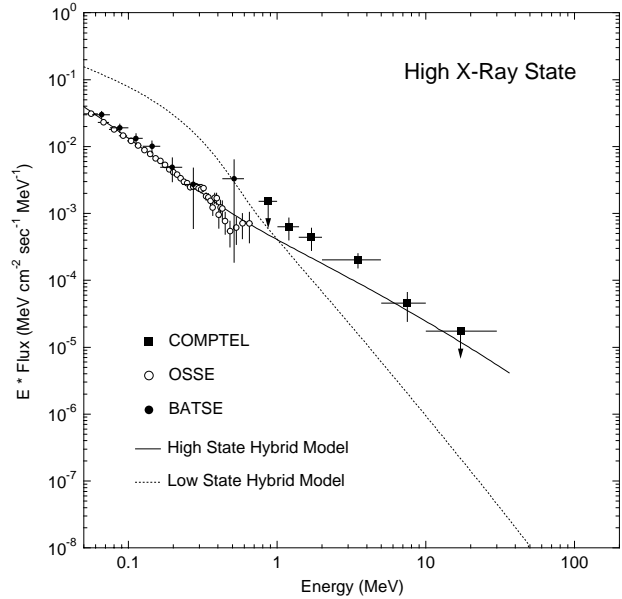

**FIGURE 3** - Broadband CGRO spectrum of the high X-ray state of Cygnus X-1 (VP 522.5). For the sake of clarity, upper limits from OSSE and BATSE are not shown, but these are consistent with the total dataset.

Whereas the low-state CGRO spectrum (Figure 2) shows the breaking type spectrum that is typical of most high-energy observations of Cyg X-1, the high-state CGRO spectrum (Figure 3) shows the power-law type spectrum that is characteristic of black hole binaries in their high X-ray state [10]. The high state power-law spectral behavior had already been reported for the 1996 transition based on observations with both BATSE [11, 13] and OSSE [7, 14]. An earlier detailed study of the broadband high state spectrum was based on data from ASCA, RXTE and CGRO/OSSE, but did not include the higher energy COMPTEL data [14]. The inclusion of the COMPTEL data in the high state spectrum provides evidence of a continuous power-law (with a photon spectral index of 2.6) extending beyond 1 MeV, up to ~10 MeV. No clear evidence for a cutoff in the power-law spectrum can be discerned from these data.

## SPECTRAL MODELING

More detailed physical modeling of the spectrum has been attemped using a hybrid thermal/non-thermal Comptonization model, which describes the photon spectrum resulting from the Compton scattering of low energy photons off a hybrid electron distribution [15]. The total electron distribution includes both a thermal (Maxwellian) distribution plus a non-thermal (power-law) distribution at higher energies. The thermal electron component is defined by a characteristic temperature ($kT_e$). The non-thermal electron component is characterized as a power law with spectral slope $p_e$ extending from an electron Lorentz factor $\gamma_{min}$ (where the Maxwellian transforms to the power-law tail) up to a Lorentz factor of $\gamma_{max}$. The electron population is assumed to reside in an accretion disk corona with an optical depth of $\tau$. This particular model is useful in that it permits a quantitative description of the underlying electron distribution based on the observed photon spectrum. We have found that this model can provide very good fits to both the low- and high-state spectra, as can be seen in Figures 2 and 3. However, with the CGRO data alone, we find it difficult to constrain many of the model parameters. In our earlier analysis of the low-state spectrum, we used an energy threshold of 200 keV [4]. Unfortunately, such a high threshold leaves the fitting process very insensitive to some of the most important model parameters, such as $kT_e$ and $\gamma_{min}$.

We are continuing the analysis of these data in an effort to provide tighter constraints on the model parameters, especially those (such as $p_e$ and $\gamma_{max}$) that are most influenced by the high-energy data. The latest analysis uses a lower threshold energy of 50 keV and (unlike those fits shown in Figures 2 and 3) also allows for independent detector normalizations. In addition, we may also incorporate data from lower energy experiments (such as RXTE and SAX). The results of this on-going analysis will be reported elsewhere [16].

## ACKNOWLEDGMENTS

The COMPTEL project is supported by NASA under contract NAS5-26645, by the Deutsche Agentur für Raumfahrtgelenheiten (DARA) under grant 50 QV90968 and by the Netherlands Organization for Scientific Research NWO. This work was also supported by NASA grant NAG5-7745.